\documentclass[aps, prl,twocolumn,reprint,superscriptaddress]{revtex4-1}
\usepackage{comment}
\usepackage{graphicx}
\usepackage{amsmath}
\usepackage{setspace}
\usepackage{braket}
\usepackage{mathtools}

\usepackage{natbib,mathrsfs,accents}
\usepackage[dvipsnames]{xcolor}
\definecolor{myblue}{named}{MidnightBlue}
\definecolor{mygreen}{RGB}{0,120,0}
\usepackage{siunitx}

\usepackage[caption=false]{subfig}
\usepackage{float}

\usepackage{url}

\begin{document}


\title{Proposal for a long-lived quantum memory using matter-wave optics with Bose-Einstein condensates in microgravity}

\author{Elisa Da Ros$^\mathsection$}
\affiliation{Institut f\"{u}r Physik and IRIS, Humboldt-Universit\"{a}t zu Berlin, Newtonstr. 15, Berlin 12489, Germany}

\author{Simon Kanthak$^\mathsection$}
\affiliation{Institut f\"{u}r Physik and IRIS, Humboldt-Universit\"{a}t zu Berlin, Newtonstr. 15, Berlin 12489, Germany}

\author{Erhan Sağlamyürek}
\email[]{Currently at Lawrence Berkeley National Lab, Berkeley, California, 94720, USA and Department of Physics,  University of California, Berkeley, California, 94720, USA}
\affiliation{Department of Physics and Astronomy, University of Calgary, Calgary, Alberta T2N 1N4, Canada}
\affiliation{Department of Physics, University of Alberta, Edmonton, Alberta T6G 2E1, Canada}

\author{Mustafa Gündoğan}
\email[]{mustafa.guendogan@physik.hu-berlin.de}
\affiliation{Institut f\"{u}r Physik and IRIS, Humboldt-Universit\"{a}t zu Berlin, Newtonstr. 15, Berlin 12489, Germany}

\author{Markus Krutzik} 
\email[]{markus.krutzik@physik.hu-berlin.de}
\affiliation{Institut f\"{u}r Physik and IRIS, Humboldt-Universit\"{a}t zu Berlin, Newtonstr. 15, Berlin 12489, Germany}
\affiliation{Ferdinand-Braun-Institut (FBH), Gustav-Kirchoff-Str.4, 12489 Berlin}

\begin{abstract}
Bose-Einstein condensates are a promising platform for optical quantum memories, but suffer from several decoherence mechanisms, leading to short memory lifetimes. While some of these decoherence effects can be mitigated by conventional methods, density dependent atom-atom collisions ultimately set the upper limit of quantum memory lifetime to s-timescales in trapped Bose-Einstein condensates. We propose a new quantum memory technique that utilizes microgravity as a resource to minimize such density-dependent effects. We show that by using optical atom lenses to collimate and refocus the freely expanding atomic ensembles, in an ideal environment, the expected memory lifetime is only limited by the quality of the background vacuum. We anticipate that this method can be experimentally demonstrated in Earth-bound microgravity platforms or space missions, eventually leading to storage times of minutes and unprecedented time-bandwidth products of {$10^{10}$}. 

\end{abstract}


\maketitle

Optical quantum memories (QMs) are devices that can faithfully and reversibly store and recall the quantum states of light. They are required in many applications in quantum information science such as long-distance quantum communications~\cite{Sangouard2011, Heshami2016}, deterministic generation of multiphoton states~\cite{Nunn2013} and quantum computation~\cite{Gouzien2021}. A recent idea is to deploy QMs in space in order to enable globe-spanning quantum networks~\cite{Gundogan2021, Liorni2021, Sidhu2021, Wallnofer2022}, ultra-long baseline Bell experiments~\cite{Cao2018, Mazzarella2021, Gundogan2021b, Lu2022} and probing the interplay between gravity and quantum physics~\cite{Barzel2022} for which a storage time, $\tau_{\mathrm{mem}}$, of around $\sim$\SI{1}{\second} is needed. Several atomic systems have been proven useful for such reversible mapping between light and matter qubits. These include single defects in diamond~\cite{Fuchs2011, Sukachev2017, Bradley2019}, rare-earth ion doped crystals~\cite{Ledingham2012, Jobez2015, Gundogan2015}, trapped ions~\cite{Langer2005, Harty2014, Wang2021}, single trapped atoms~\cite{Specht2011, Langenfeld2021}, and warm~\cite{Cho2016, Wolters2017, Katz2018, Kaczmarek2018} and cold~\cite{Riedl2012, Wang2019, Bao2012, Heller2020, Saglamyurek2021} atomic gases. Among these platforms, cold-atomic gases have recently been deployed in space for a number of experiments: optical atomic clocks~\cite{Liu2018}; the first Bose-Einstein condensate (BEC) on board a sounding rocket~\cite{Becker2018} and the International Space Station (ISS)~\cite{Aveline2020}. In addition to these, missions using cold atoms in space are being envisioned~\cite{Roadmap2022} for gravity and dark matter exploration~\cite{Tino2019, ElNeaj2020}, and currently in development for ultra-cold atom research including atom interferometry~\cite{Beccal2021} and advanced atomic clocks on board the ISS~\cite{Cacciapuoti2020}. Cold atom based QMs would share the same technical infrastructure with these experiments.

\begin{figure*}[t!]
\begin{center}

\includegraphics[width=1.8\columnwidth]{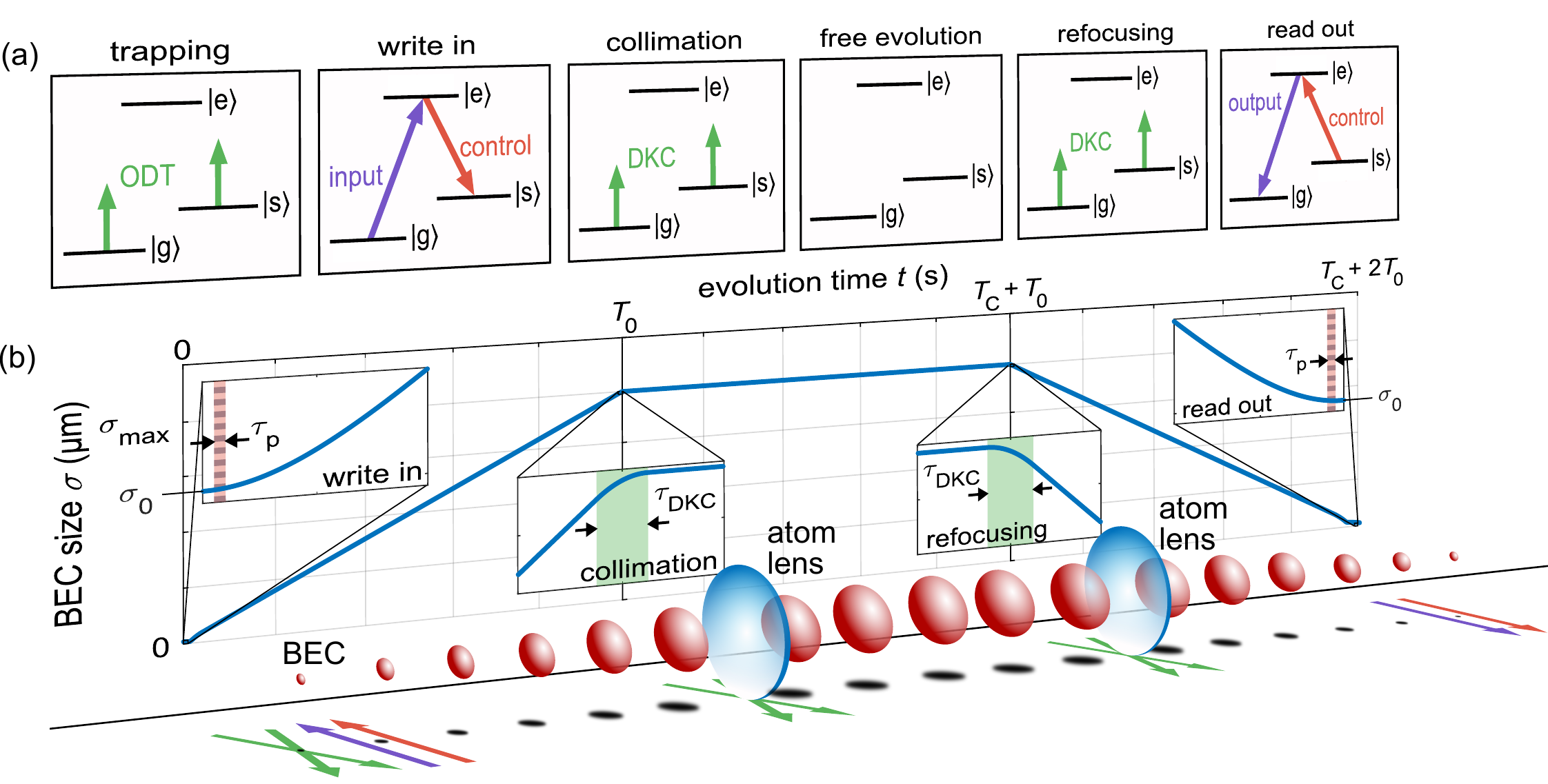}\label{fig:1a:protocol}
\caption{Protocol for a long-lived quantum memory utilizing interaction-driven expansion and delta-kick collimation (DKC) of a Bose-Einstein condensate (BEC) in microgravity. a) $\Lambda$-type three-level structure together with the employed light fields during b) different stages of the size evolution of the BEC. The quantum state of single photon pulses is imprinted into an internal excitation of a BEC shortly after its release from an optical dipole trap (ODT). Brief exposure of the BEC by two consecutive optical lensing potentials allows to stop and subsequently revert the interaction-driven expansion via DKC. This protocol allows for transition between the complementary density regimes needed for an efficient write-in and read-out of the memory at high optical depths (ODs) and large coherence times for long-time storage at low atomic densities, respectively. }
\label{fig:comparison_single_memory}
\end{center}
\end{figure*}

A BEC platform has unique advantages over cold atoms (obeying a thermal distribution) for optical QMs due to the inhibition of thermal motion (allowing long memory lifetime) and its high atomic density (leading to efficient operation). However, condensates are still affected by several decoherence mechanisms. Among these, decoherence due to magnetic field inhomogeneities~\cite{Riedl2012, Saglamyurek2021} can be mitigated by employing rephasing protocols based on dynamical decoupling~\cite{Dudin2013} and those caused by AC Stark shifts that are due to inhomogeneous optical trapping beams can be prevented by employing magic wavelength techniques~\cite{Lundblad2010, Dudin2010a}. On the other hand, losses due to atom-atom collisions are usually not reversible, and cannot be mitigated by such measures. The collisions of cold atoms with the background gas (i.e. 1-body collisions) can be controlled only with the vacuum quality, while the collision rates between 2 or 3 atoms within the cold ensemble (i.e. 2-body and 3-body collision) increase with increasing atom density. 
These processes become relevant beyond storage times of $\sim\SI{1}{ms}$. However, a maximum storage time of around $\sim\SI{1}{s}$ has been observed with bright pulses in a Sodium BEC by tuning the atom-atom collision cross sections via external magnetic fields~\cite{Zhang2009}.  

In this work, we propose a novel quantum storage scheme that exploits matter-wave optics to tune the density of the atomic ensemble to minimize the effects of density-dependent collisions. This is achieved by letting the condensate expand after writing the quantum state of incoming photons into an internal state of the atoms in the condensate, which is followed by employing the delta kick collimation (DKC) technique~\cite{Ammann1997,Myrskog2000,Kovachy2015_PRL,Deppner2021_PRL}, first to collimate and then to refocus the BEC for efficient read-out of the stored excitation. This protocol is carried out in a microgravity environment, which prevents the fall of the centre of mass without the need for any types of inhomogeneous field to levitate the atoms. We show that this technique would allow storage times that are orders of magnitude beyond what is possible in ground-based experiments and, in fact, only limited by the quality of the background vacuum. We expect our protocol to reach a few minutes of storage time with the state-of-the-art background vacuum values~\cite{Hogan2011, Nirrengarten2006}. 

We assume a pure BEC initially trapped in an optical dipole trap (ODT), as illustrated in Fig.~1. To circumvent decoherence due to AC-Stark shifts, the quantum state of single photon pulses is imprinted within the BEC only shortly after its release from the trap. Timing of the write pulse is set to mode-match the light intensity and atomic density distributions with negligible reduction in optical depth (OD). During free expansion, the internal energy is converted into kinetic energy, yielding a reduction in the density and therefore in the 2-body collisions. After a set time $T_0$, the BEC is exposed to a tailored, optical potential for a short duration of $\tau_\text{DKC}$. This way, the BEC experiences a delta-kick, which acts as an optical atom lens~\cite{Kovachy2015_PRL,Luan2018,Kanthak2021,Gochnauer2021}, resulting in a narrow momentum distribution. After a chosen collimation time $T_\text{C}$, a second DKC pulse is applied to refocus the ensemble. At this point ($T_\text{C} +  2\,T_0$) it is possible to faithfully recall the stored quantum information at the original higher OD. Our protocol thus allows to transition between the complementary density regimes needed for an efficient write-in and read-out at high ODs and coherent storage in a dilute quantum gas by exploiting the mean-field driven expansion of a self-interacting BEC. Given by the point-like source characteristics and single-mode properties of the BEC, the dispersion of the ensemble can be shaped after release from the trapping potential by DKC to nearly stop and finally revert the expansion. 

The quantum memory itself is based on a $\Lambda$-type three-level system as represented in Fig.~1. The states $\ket{g}$ and $\ket{s}$ represent the ground states of the hyperfine structure of the $^{87}$Rb D$_1$ line, $\ket{5^2S_{1/2}, F=1}$ and $\ket{5^2S_{1/2}, F=2}$, respectively, while  $\ket{e}$  is the excited state $\ket{5^2P_{1/2}, F=1}$. Collinear probe and control beams address the $\ket{g}\longleftrightarrow\ket{e}$ and $\ket{s}\longleftrightarrow\ket{e}$ transitions, respectively. The use of collinear beams~\cite{Namazi2017} ensures both the optimal spin storage and phase-matching condition which in turn eliminates decoherence due to recoil collisions~\cite{Saglamyurek2021}.

Although ensemble-based memories generally follow similar considerations~\cite{Gorshkov2007}, we choose to incorporate the Autler–Townes splitting (ATS) method~\cite{Saglamyurek2018, Saglamyurek2021} into our approach as it requires lower OD and control power for efficient storage of broadband pulses compared to other memory protocols implemented in cold-atom systems, such as electromagnetically induced transparency~\cite{Geng2014}, which makes it more attractive for applications in quantum information science. Furthermore, lower requisites on these properties make the ATS protocol more robust against four-wave mixing noise~\cite{Saglamyurek2021}, which is another important feature for practical applications.

\noindent

We predict the dynamics of the BEC through a variational ansatz to numerically solve the time-dependent Gross–Pitaevskii equation

\begin{equation}
\label{eq:GPE}
\text{i}\hbar \frac{\partial}{\partial t} \psi(\vec{r},t) =\\ \left[-\frac{\hbar^2}{2m}\nabla^2+V_\text{DKC}(\vec{r},t)+U_0\rho(\vec{r},t)\right] \psi(\vec{r},t)
\end{equation}
\noindent
 following a scaling approach~\cite{Garcia1997_PRA, Garcia1996_PRL}, where $V_\text{DKC}(\vec{r},t)$ represents the lensing potential, $U_0 = 4\pi \hbar^2 \text{Re}(a_\text{sc}) N_0 /m$ characterizes the interaction and is defined by the real part of the s-wave scattering length $a_\text{sc}$~\cite{Harber2002} for a ground state BEC with $N_0$ atoms of mass $m$.

As trial functions, we simply utilize a Gaussian ansatz for the atomic wavefunction $\psi(\vec{r},t)$  with corresponding spatial density distribution $\rho(t)$ expressed as 

\begin{equation}
\label{eq:WF}
\rho(t) = \left|\psi(t)\right|^2= \underbrace{\frac{N_0}{(2\pi)^3} \prod_{\xi\in \{x,y,z\}} \sigma_\xi^{-1}(t)}_{\text{\normalsize$=\rho_0(t)$}}\exp\left(-\frac{\xi^2}{2\sigma_\xi^2(t)}\right),
\end{equation}
\noindent
where $\rho_0(t)$ represents the peak density. The standard deviation $\sigma_\xi(t)$ of the atomic density can then be related to the Thomas-Fermi radius $\sigma_\xi = R_\xi/ \sqrt{7}$, knowing the parabolic shape of a BEC~\cite{Corgier2020_NJP}.

The lensing potential is well described by the harmonic approximation $V_\text{DKC}(\vec{r},t)= 1/2 m \sum_\xi \omega_\xi^2\xi^2$ if the BEC is located close to the center of the optical trap and if the characteristic size of the generating beams is $w_0\gg \sigma_\xi$. Any anharmonicity of the lensing potential would cause lens aberrations and will ultimately limit the achievable storage times and efficiency of the information read-out due to the attainable collimation times~\cite{Deppner2021_PRL} and minimum sizes during refocusing~\cite{Kovachy2015_PRL}, respectively.

For our case study, we initialize an isotropic BEC of size $\sigma = \SI{3}{\micro\meter}$ with $N_0 = 1\times10^5$ $^{87}$Rb atoms. After $T_0=\SI{1}{\second}$ of free expansion, we apply a DKC potential with trap frequency $\omega=2\pi\cdot\SI{2.25}{\hertz}$ for a symmetrically centered box pulse of duration $\tau_\text{DKC}=\SI{5}{\milli\second}$. After further evolution for $T_\text{C}$, the BEC is exposed again to the same DKC potential. The trap frequency is chosen to mode-match the final and initial wavefunction. The total storage time of the memory in this protocol is then approximately $\tau_\text{mem}\approx T_\text{C} +  2\,T_0 $.

In general, illumination of the ensemble with an inhomogeneous light field, as needed for optical DKC, can affect the system coherence. Assuming a crossed beam ODT at \SI{1064}{\nano\meter} as origin of the lensing potential, we calculate the differental AC-Stark shift of the two ground states of $^{87}$Rb to be $\delta \nu_\text{AC}\ll \tau_\text{DKC}^{-1}$, granting a negligible decoherence during the DKC pulses.

\begin{figure}[H]
\centering
\includegraphics[width=1\columnwidth, trim=0.92cm 0.75cm 9cm 6cm, clip]{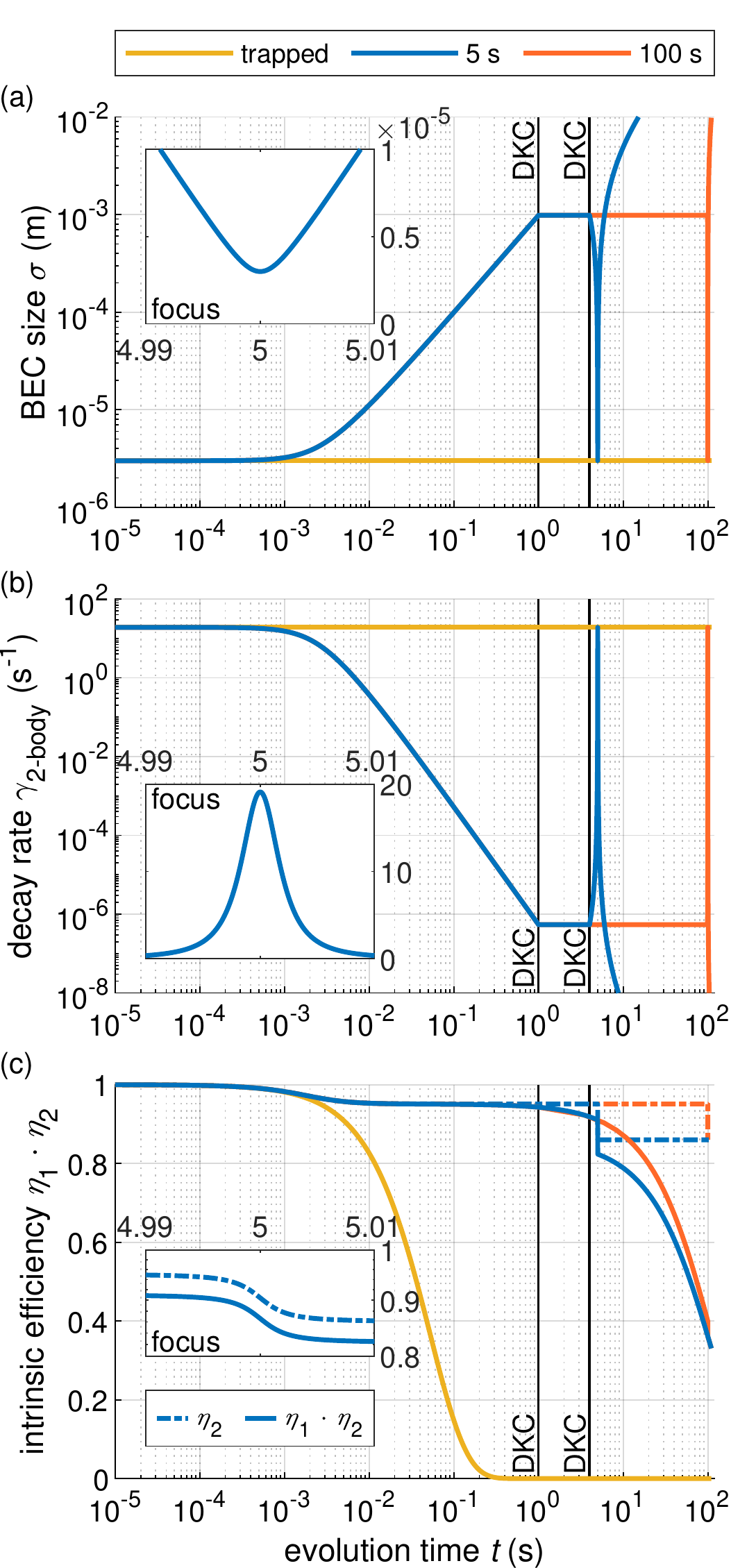}
\caption{BEC dynamics and intrinsic memory efficiency. Time evolution of a) the BEC size $\sigma$, b) corresponding decay rates $\gamma_{\text{2-body}}$ and its effect on c) the intrinsic memory efficiency (i.e., not including write-in and read-out efficiencies) due to only 2-body (dashed lines) and combination of 2- and 1-body collisions (solid lines). We compare our memory protocol with storage times $\tau_\text{mem}\approx\SI{5}{\second}$~(blue) and $\tau_\text{mem}\approx\SI{100}{\second}$~(orange) together with the trapped case~(yellow). The inlets show a zoom onto the respective changes during the focus of the BEC for the case $\tau_\text{mem}\approx\SI{5}{\second}$, in linear scaling. The black vertical lines indicate the timing of the DKC pulses to collimate and refocus the BEC for the case with $\tau_\text{mem}\approx\SI{5}{\second}$.}
\label{fig:comparison_evolution}
\end{figure}

Assuming that the untrapped ensemble is shielded from any magnetic field inhomogenities, the overall efficiency of photon retrieval can then be expressed as a function of the atomic density distribution and peak density as
\begin{equation}
\label{eq:overall_eff}
\eta_{\text{tot}}(t)= \eta_{\text{DKC}}^2\cdot \eta_{\text{1}}(t) \cdot \eta_{\text{2}}(\rho_0(t)) \cdot \eta_{\text{ATS}}(\rho(t)), 
\end{equation}
\noindent
where $\eta_{\text{DKC}}$ indicates the efficiency of the DKC procedure, $\eta_{\text{1}}(t)$ and $\eta_{\text{2}}(\rho_0(t))$ are associated with the 1-body and 2-body collisions, respectively, and finally the factor $\eta_{\text{ATS}}(\rho(t))$ represents the combined efficiencies of the ATS write-in and read-out steps. 

For the purpose of this work, $\eta_{\text{DKC}}$ is set to 1. This corresponds to the assumption of ideal harmonic potentials for the implementation of the DKC procedures, and thus neglects possible lens aberrations that would affect the atom distribution \cite{Kovachy2015_PRL}. This ideal case  might be accomplished in experiments utilizing time-averaged optical potentials~\cite{Albers2022}.

The efficiency associated with the atom losses due to the collisions with the background gas within the vacuum chamber, $\eta_{\text{1}}(t)$, follows an exponential decay. In~\cite{Nirrengarten2006}, Nirrengarten et al. achieve a lifetime of about $\SI{115}{s}$ associated with a background pressure of $\SI{3e-11}{\milli\Bar{}}$. We set this value as the 1-body collision lifetime $\tau_{1}$ used within this work. 
The term $\eta_{\text{2}}(\rho_0(t))$ expresses the exponential decay of the memory efficiency due to two-body collisions between cold Rb atoms. For the states and densities taken into account in this work, the contribution of the 3-body collisions is negligible compared to the other collisional losses~\cite{Burt1997}. 

\begin{figure}[t!]
\begin{center}
\includegraphics[width=1\columnwidth,trim=0.92cm 0.75cm 9cm 20.5cm, clip]{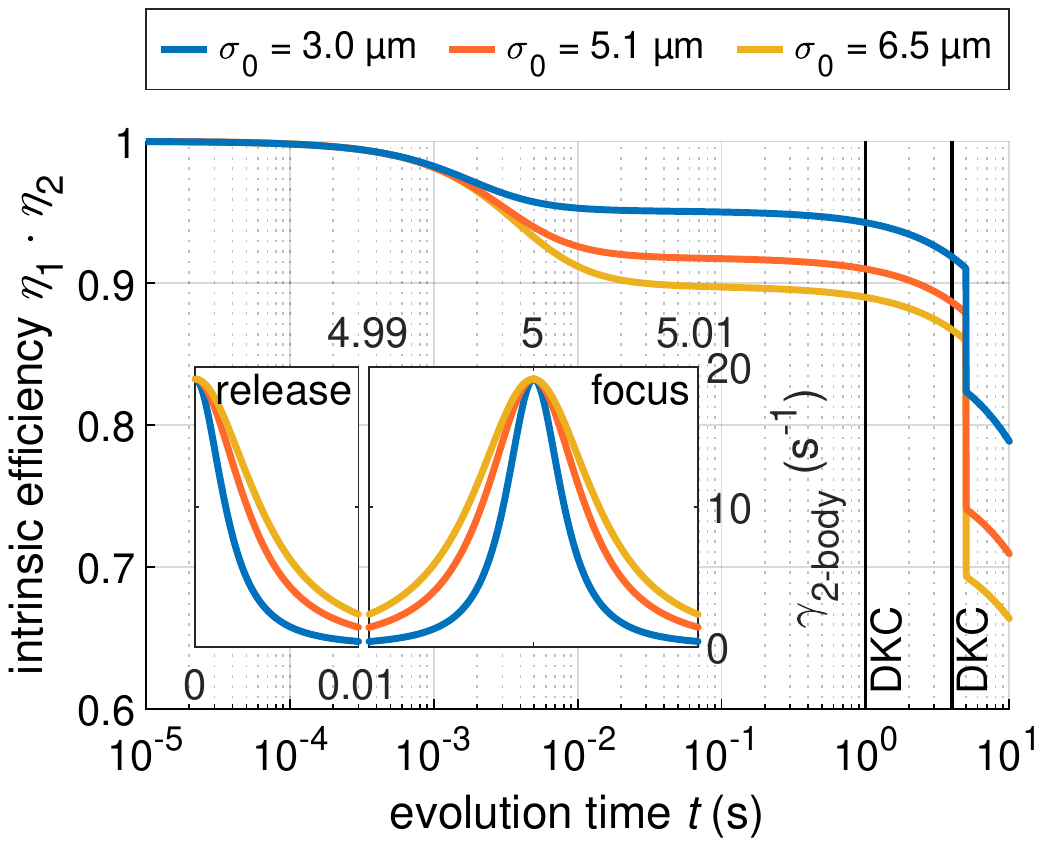}
\caption{Intrinsic memory efficiency for different initial BEC sizes $\sigma_0$ and common initial peak density $\rho_\mathrm{0}(\text{0})=\SI{2.3e14}{\per\cubic\centi\metre}$. The black lines indicate the timing of the DKC pulses to collimate and refocus the BEC. The inlet shows the evolution of the 2-body decay rate shortly after release and during refocusing, in linear scaling.}
\label{fig:Intrinsic_Eff_vs_trap_freq}
\end{center}
\end{figure}

Figure~\ref{fig:comparison_evolution}$\,$(a) shows the comparison of the time evolution of the condensate size for two different collimation times (blue and orange lines) with the scenario in which the BEC is not released from the trap at all (yellow line). The color code is the same for the other panels in the figure. In both expanding cases, the condensate peak density drops from its initial value $\rho_\mathrm{0}(\text{0})=\SI{2.3e14}{\per\cubic\centi\metre}$ to  $\rho_\mathrm{0}(T_\text{0}=\SI{1}{\second})=\SI{7e6}{\per\cubic\centi\metre}$ at the time of the first DKC pulse. After a duration of $T_\text{C}=\SI{3}{\second}$ (blue) or $T_\text{C}=\SI{98}{\second}$ (orange), the second DKC pulse is applied, which causes the BEC to refocus and reach again its minimum size at $\tau_{\text{mem}} \approx T_\text{C}+2\,T_\text{0}$, when the memory is read out. The corresponding storage times are $\tau_\text{mem}\approx\SI{5}{\second}$ (blue) or  $\tau_\text{mem}\approx\SI{100}{\second}$ (orange).  Upon refocusing, the initial density of the ensemble is fully recovered and, subsequently, the BEC keeps expanding again (as shown by the blue line after $\SI{5}{\second}$). The inset shows the detailed evolution of the BEC size around the focus for the case with $T_\text{C}=\SI{3}{\second}$.


The time evolutions of peak density corresponding to the simulated sizes of Fig. \ref{fig:comparison_evolution}~(a) allow to calculate the time-dependent 2-body collisions decay rate $\gamma_{\text{2-body}}(t)$ as~\cite{Egorov2012, Saglamyurek2021}
\begin{equation}
\label{eq:coll_rate}
\gamma_{\text{2-body}}(t)=\frac{4~h \text{Im}(a_{\text{sc}}) \rho_0(t) }{m}, 
\end{equation}
\noindent
where $\text{Im}(a_{\text{sc}})$ is the imaginary part of the s-wave scattering length. Figure~\ref{fig:comparison_evolution}$\,$(b) shows a reduction of more than 7 orders of magnitude in the 2-body decay rate during $T_0$, compared to the trapped case with constant peak density. The inset highlights again the behaviour around the focus for the case with $T_\text{C}=\SI{3}{\second}$.

With this time-dependent decay rate, the intrinsic efficiency due to 2-body collisions is found upon integration over time as 
\begin{equation}
\label{eq:coll_eff}
\eta_{\text{2}}(\rho_0(t)) =\exp\left(-\int_0^{t} \frac{\rho_0(t')}{\kappa}\mathrm{d} t'\right)
\end{equation}
where $\kappa = m/ \left( 4 h \text{Im}(a_{\text{sc}})\right)$. 

The intrinsic memory efficiency, $\eta_{\text{1}}(t)\cdot~\eta_{\text{2}}(\rho_0(t))$, is plotted in Fig.~\ref{fig:comparison_evolution}$\,$(c).
The solid lines include the effects of 1- and 2-body collisions whereas the dashed lines show only the effects of 2-body collisions. It is evident from the figure, that the constant high density of the trapped case yields a lifetime limit to around $\SI{100}{\ms}$ by 2-body collisions. On the contrary, the lifetime can be extended for the expanding and refocused cases, proving that the memory decay rate can be tuned by varying the BEC density via DKC pulses. The spike that occurs for $\gamma_{\text{2-body}}$ after the second DKC pulse (see Fig.~\ref{fig:comparison_evolution}$\,$(b)), due to the increased density during refocusing, results in a sudden drop in intrinsic efficiency, as seen in the inset of Fig.~\ref{fig:comparison_evolution}$\,$(c). Nonetheless, this drop is not particularly detrimental: from 0.95 to 0.86. The inset focuses on the comparison between the effects of 1- and 2-body collisions, $\eta_{\text{1}}\cdot~\eta_{\text{2}}$, and the single 2-body  contribution, $\eta_{\text{2}}$, around the focus, for the case $\tau_\text{mem}\approx\SI{5}{\second}$. 
It highlights how the main loss contribution on the short timescales is given by 2-body collisions (e.g. the contribution of $\gamma_{\text{1-body}}$ is limited to about $4\%$). The 1-body collisions become, instead, the dominant loss mechanism at long storage times. For instance for $\tau_{\text{mem}}\approx\SI{100}{\second}$, the intrinsic efficiency due to $\gamma_{\text{2-body}}(t)$ is still around 0.90, whereas the collisions with the background gas brings it down to 0.38. 

The initial atom density distribution of the BEC determines the expansion rate of the ensemble, which consequently affects the achievable intrinsic memory efficiency. Figure~\ref{fig:Intrinsic_Eff_vs_trap_freq} shows the trends of the intrinsic memory efficiency for a set of three isotropic BECs with different initial sizes $\sigma_0$. All cases have, though, a common initial peak density $\rho_\mathrm{0}(\text{0})=\SI{2.3e14}{\per\cubic\centi\metre}$ to grant a common initial $\gamma_{\text{2-body}}(0)$. The ensembles expand for $\SI{1}{\second}$, until collimation, and are then refocused after $T_\text{C}=\SI{3}{\second}$. For smaller initial size, the ensemble expands faster, yielding lower accumulated losses from the 2-body collisions and therefore granting higher intrinsic efficiencies. The inset displays the trends of the 2-body decay rate for the three cases during the initial expansion phase and around the focus.

\begin{figure}[t!]
\begin{center}
\includegraphics[width=1\columnwidth,trim=0.92cm 0.75cm 9cm 20.5cm, clip]{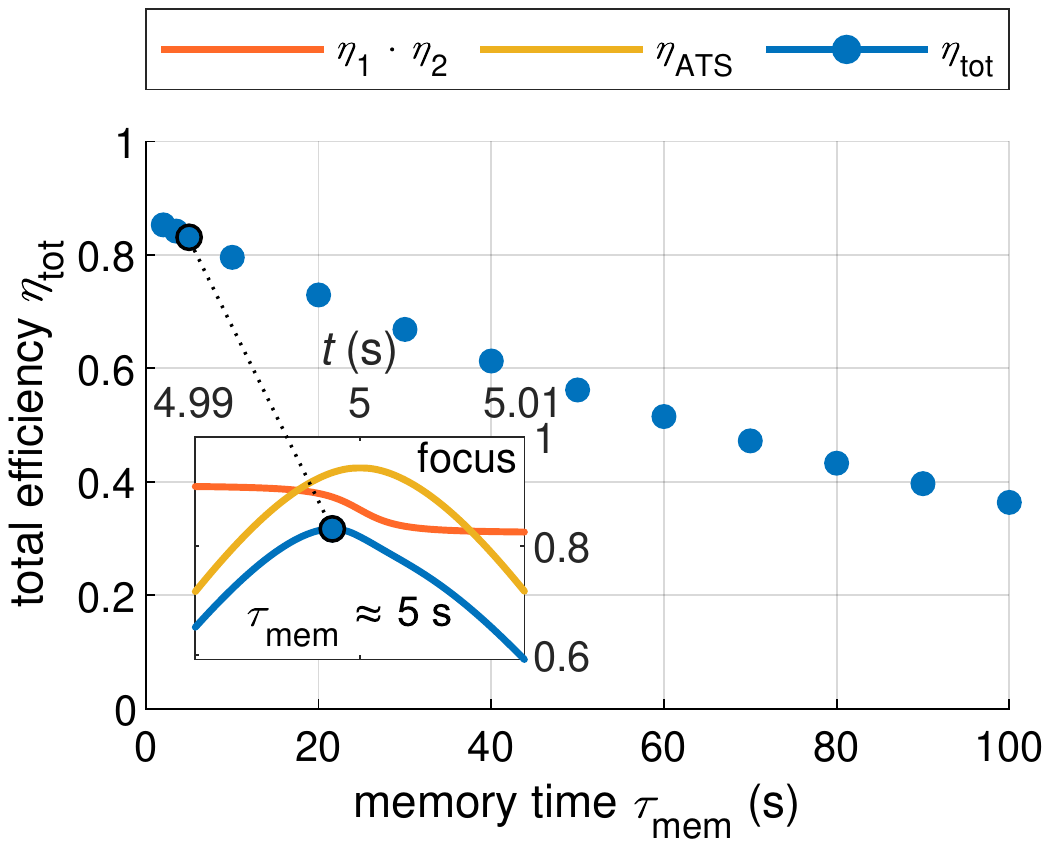}
\caption{Calculated overall efficiency as a function of the storage time in BECs of initial size $\sigma_0= \SI{3}{\micro\meter}$. For each data point the readout is performed after refocusing, i.e. at high density. The inset shows the comparison between the overall retrieval efficiency $\eta_\text{tot}$(blue), the efficiency factor $\eta_\text{ATS}$ and the intrinsic memory efficiency $\eta_1\cdot\eta_2$ as a function of time for $T_{\text{C}} = \SI{3}{\second}$ (i.e., $\tau_{\text{mem}}\approx\SI{5}{\second}$) during the focus of the BEC.}
\label{fig:Efficiency_vs_storage_time}
\end{center}
\end{figure}

In order to calculate the overall efficiency, $\eta_{\text{tot}}$, we estimate the optimal write-in and read-out efficiency associated with an ATS protocol (in this case in backward retrieval) as~\cite{Rastogi2019}
\begin{equation}
\label{eq:eta_ATS }
\eta_\text{ATS}(\rho(t)) \approx \left(1-e^{-d(t)/(2 F)}\right)^2e^{-d(t)/(2 F)},
\end{equation}
\noindent
where $F\approx 2\pi B / \Gamma$ is the ‘ATS factor’ depending on the bandwidth $B$ and the optical transition linewidth $\Gamma$. The effective OD, $d(t)$, results from integration along the probe beam propagation of the spatial overlap between the Thomas-Fermi density distribution of the ensemble and intensity profile of the probe beam~\cite{Saglamyurek2021}. 
In this work, we consider a probe beam with a waist of $\SI{1}{\micro\meter}$, which is smaller than the radius of the BEC $\sigma(t)$, and a Gaussian temporal profile with full width at half maximum $\tau_{p}=\SI{2.4}{\nano\second}$. The associated bandwidth, $B=\SI{180}{\mega\hertz}$, satisfies the condition for optimal ATS efficiency associated with the condensate OD and the optical transition linewidth $\Gamma$~\cite{Rastogi2019}. 

The total efficiency for a series of on-demand read-out cases with different storage times and initial size $\sigma_0=\SI{3}{\micro\meter}$ is calculated through Eq.~\ref{eq:overall_eff} and presented in Fig.~\ref{fig:Efficiency_vs_storage_time}. The inset highlights separately the contributions of the ATS efficiency factor $\eta_{\text{ATS}}$ (yellow) and of the intrinsic efficiency $\eta_{\text{1}}\cdot\eta_{\text{2}}$ (orange) to the total efficiency $\eta_\text{tot}$ (blue) for the case with $\SI{5}{\second}$ storage time. It is worth noting how the increased 2-body decay rate that derives from the enhanced density in the focus leads to a shift in the efficiency peaks: the peak in the total efficiency is registered consistently $\delta\tau\approx\SI{2}{\milli\second}$ before the peak density is achieved. The magnitude of this effect depends on the expansion rate of the BEC: a slower expansion of the ensemble generally corresponds to a larger value of $\delta\tau$. In the case with the slowest expansion considered in Fig.~\ref{fig:Intrinsic_Eff_vs_trap_freq}, corresponding to the initial size $\sigma_0=\SI{6.5}{\micro\meter}$ (yellow line), for instance, the shift would rise to $\delta\tau\approx\SI{7}{\milli\second}$.
Each point presented in the main figure is, thus, associated with the peak total write-in and readout efficiency after the BEC refocusing. Assuming ideal lensing potentials, overall efficiencies $>$35\% can be reached for storage times up to $\SI{100}{s}$, neglecting external sources of noise.
With a storage time of $\tau_\text{mem}\approx\SI{100}{s}$, the proposed memory will thus exhibit an unprecedented time-bandwidth product of $\sim\num{1e10}$. 

In conclusion, by focusing on density-dependent effects on the memory efficiency, we proposed a new experimental protocol that extends the storage time of BEC-based quantum memories. This method relies on matter-wave lensing to tune the density of an expanding pure BEC in microgravity. We show storage times up to $\tau_\text{mem}\approx\SI{100}{s}$, which are ultimately limited by the vacuum quality and the consequent collisions with the background gas.
Note that aberrations due to anharmonicity of the lensing potentials or possible technical field inhomogeneities are not accounted for in our simulations. Nevertheless, this method can open up the way to push the limit of cold atom based quantum memories to $\tau_\text{mem}\approx\SI{100}{s}$ storage time, which would be of interest for growing demands of long storage time memories for space operations~\cite{Cao2018, Mazzarella2021, Gundogan2021b, Lu2022}.

Furthermore, this protocol that relies on interaction-driven expansion to decrease the atomic density, and consequently the collisional losses, followed by matter-wave optics techniques to collimate and refocus the ensemble has advantages that are not restricted to the quantum memory field, but can have applications for a broader audience. 
Proof of principle tests of this technique should be readily possible within long baseline facilities \cite{Hardman2016, Asenbaum2017, Wodey2020}, microgravity platforms on ground, e.g. drop tower facilities~\cite{Deppner2021_PRL, Muntinga2013_PRL} and ultimately in space~\cite{Becker2018, Aveline2020, Beccal2021}.

\section{Acknowledgements}
\noindent This work is supported by the German Space Agency
(DLR) with funds provided by the BMWK under grant number No.~50WM2055 (OPTIMO-II)  and No.~50WM2250B (QUANTUS+). M.~G. further acknowledges funding from the European Union's Horizon 2020 research and innovation programme under the Marie Skłodowska-Curie grant agreement No.~894590.

\noindent $^\mathsection$~E.~D.~R. and S.~K. contributed equally.

\bibliography{main4}

\end{document}